\documentclass{article}

\usepackage{arxiv}

\usepackage[utf8]{inputenc}
\usepackage[T1]{fontenc}
\usepackage{hyperref}
\usepackage{url}
\usepackage{booktabs}
\usepackage{amsfonts}
\usepackage{nicefrac}
\usepackage{microtype}
\usepackage{lipsum}
\usepackage{graphicx}
\usepackage[round,authoryear]{natbib}
\usepackage{doi}
\usepackage{enumitem}

\title{Moving beyond spatial and random cross-validation in environmental modelling: a call for prediction-domain adaptive evaluation}

\author{ \href{https://orcid.org/0000-0003-0991-8646}{\includegraphics[scale=0.06]{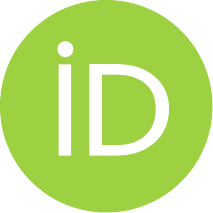}\hspace{1mm}Jan Linnenbrink*} \\
	Institute of Landscape Ecology\\
	University of Münster\\
	Münster, 48149 \\
	\texttt{jan.linnenbrink@uni-muenster.de}\\
    {*}Shared first authorship\\
	\And
	\href{http://orcid.org/0000-0002-1057-3721}{\includegraphics[scale=0.06]{orcid.pdf}\hspace{1mm}Jakub Nowosad*} \\
	Institute of Landscape Ecology\\
	University of Münster\\
	Münster, 48149 \\
	\texttt{nowosad.jakub@gmail.com} \\
    {*}Shared first authorship\\
	\AND
    \href{https://orcid.org/0000-0003-0556-0210}{\includegraphics[scale=0.06]{orcid.pdf}\hspace{1mm}Hanna Meyer} \\
	Institute of Landscape Ecology\\
	University of Münster\\
	Münster, 48149 \\
	\texttt{hanna.meyer@uni-muenster.de} \\
}

\renewcommand{\shorttitle}{Prediction-domain adaptive evaluation}

\hypersetup{
pdftitle={Prediction-domain adaptive evaluation},
pdfauthor={Jan Linnenbrink, Jakub Nowosad, Hanna Meyer},
pdfkeywords={area of applicability; data leakage; machine learning; map accuracy; model validation; spatial prediction},
}

\begin{document}
\maketitle

\begin{abstract}
	With the growing application of spatial predictive modeling in ecology, the question of how to appropriately evaluate the resulting maps has gained increasing attention. While there is consensus that map accuracy is ideally estimated using an independent probability sample of the prediction area, there is still no agreement on the most appropriate way to conduct an evaluation for the common case when such a sample is not available. 

    Cross-validation, which involves multiple train–test splits, is commonly applied not only to estimate final model accuracy but also to guide model tuning and selection. Many different spatial and non-spatial approaches to cross-validation have been proposed, and approaches in both groups have faced substantial criticism. It has been shown that random cross-validation methods are suitable when the training points are randomly distributed in the prediction area, while spatial cross-validation is better suited towards extrapolation situations. In practice, however, there is a continuum and most cases are between those two extremes.
    
    To address this gap, we advocate for a new category of cross-validation methods to account for this: \emph{prediction-domain adaptive evaluation}. Methods in this category flexibly adapt to the prediction situation, yielding most reliable estimates of map accuracy across different scenarios. To ground this perspective empirically, we reproduce a simulation study that was used in earlier research and systematically compare different evaluation methods and discuss their purpose.
    
\end{abstract}

\keywords{area of applicability \and data leakage \and machine learning \and map accuracy \and model validation \and spatial prediction}

\section{Introduction}
Spatial predictive modeling and mapping are essential in ecology and geosciences to assess species distributions, habitat suitability, and ecosystem dynamics across landscapes. The process typically requires two sets of data: response data representing the variable of interest, which are often drawn from large, opportunistic ecological databases (e.g., GBIF for biodiversity data: \citealt{GBIF2025}, WoSIS Soil Profile Database: \citealt{Batjes2017}, Fluxnet for gas exchange measurements: \citealt{Pastorello2020}, or sPlotOpen: \citealt{Sabatini2021}, a global dataset of vegetation surveys to mention just a few), and predictor variables, commonly derived from remote sensing products or climate models. Machine learning methods are then frequently used to train models on these data, as they are well suited to capture complex, non-linear relationships between environmental predictors and responses. Once trained, the model is applied across the entire spatial domain to generate predictions that are visualized as maps. Models may encounter different prediction scenarios, ranging from spatial interpolation (i.e., the whole prediction area is well covered by training points) to spatial extrapolation (i.e., the prediction area is not covered at all). To assess the quality of the derived map, especially as a baseline for decision-making, an evaluation is performed where the predictions are compared to data points that have not been used during model training.
The question of how the evaluation should be performed, however, is a topic that is currently the subject of ongoing discussion. Established approaches, such as design-based inference relying on probability samples of the prediction area with known inclusion probabilities, are frequently infeasible, particularly in the absence of a formal sampling design when using large opportunistic ecological databases.

As outlined in standard textbooks on statistical learning \citep{Hastie2009, Kuhn2013}, it is recommended that, if the dataset is large enough, evaluation is conducted by splitting the available data into training, validation and test data, where validation data are used for model selection and test data are used to evaluate the final model predictions. An alternative evaluation strategy that allows the entire dataset to be used for model training is cross-validation. Cross-validation involves splitting the data into multiple folds and iteratively training models, each time using one fold as the test set and the remaining folds for training. The resulting performance statistics are then communicated to describe the quality of the predictions for the study area (i.e., the map accuracy). Cross-validation has proven especially useful during model selection, including hyperparameter tuning \citep{Schratz2019, Huang2024} and predictor selection \citep{Meyer2018, Meyer2019}. However, if cross-validation is used both for model selection and for evaluating the final model, it can lead to data leakage \citep{Hastie2009, Kuhn2013}, hence using a separate test set is favourable. Still, when the training dataset is small, as is often the case in ecology and geosciences, relying on cross-validation performance to estimate final map accuracy may represent a compromise between obtaining an unbiased evaluation and ensuring adequate model training. Consequently, in the majority of studies conducted in these fields, the model selection -- and often even the final map evaluation -- is solely based on cross-validation \citep[e.g.,][]{Hoogen2019, Bastin2019, Cai2023}.
Both strategies -- using a distinct test set, or applying cross-validation -- rely on splitting the data into training and test data. In this context, the way the data are split becomes critical. In the following, we first discuss existing approaches to data splitting before proposing a new strategy for data partitioning that, we argue, provides a consolidated perspective on existing strategies.

Traditionally, and coming from an assumption of data being independent and identically distributed (iid), this data splitting was done in a random way (i.e., a single random subset used as test data or random cross-validation). However, since the distribution of spatial ecological data can rarely be regarded as iid \citep[see Figure 1 in][]{Stehman2000, Hughes2021, MeyerPebesma2022}, \emph{spatial cross-validation} approaches have been suggested.
Spatial cross-validation refers to a group of cross-validation strategies where spatial resampling is applied to prevent test points from being located close to the training points. These spatial resampling methods can be similarly applied to split the data into a single training and test set. A common approach to spatial cross-validation is spatial block cross-validation where the spatial domain is divided into several blocks, and then iteratively, all data falling into one of the spatial blocks are held back for evaluation \citep{Roberts2017, Valavi2018}. Several other spatial cross-validation strategies have been suggested (see section \ref{sec:independence}), all of them striving for independence between training and test data. Spatial cross-validation has been adopted and is currently used in many ecological fields such as species distribution modeling \citep[e.g.][]{Collette2026, Azevedo2026} or vegetation modelling \citep[e.g.][]{Lusk2026}. Notably, and as discussed by \citet{Wadoux2021}, spatial cross-validation is still poorly defined, encapsulating various approaches with numerous assumptions and parameters being proposed. 

Previous studies have shown that our perception of prediction performance can vary considerably depending on the evaluation strategy used. For example, in the case study presented in \citet{Ploton2020}, the $R^2$ for predicting aboveground forest biomass in central Africa varied between 0.14 and 0.53, depending on the evaluation strategy. \citet{Ludwig2023} demonstrated large differences between the estimated map accuracy resulting from random and a spatial cross-validation for global maps of nematode abundances ($R^2$ of 0.46 vs. 0.19) and specific leaf area ($R^2$ of  0.62 vs. 0.30). Due to the large differences in estimated map quality, a discussion has emerged regarding which of the many resampling strategies should be used in the context of machine learning for spatial environmental mapping.
As highlighted by \citet{MeyerPebesma2022}, there is an ongoing debate in which \citet{Roberts2017} and \citet{Ploton2020} criticise commonly used random cross-validation for ignoring spatial autocorrelation, which may lead to overly optimistic performance estimates. \citet{Wadoux2021} disagree with a paper provocatively entitled "Spatial cross-validation is not the right way to evaluate map accuracy". Instead, they advocate that probability sampling and design-based inference should be used and, if a probability sample is not available, random cross-validation might be an alternative. Spatial cross-validation, they argue, should be completely avoided, and this led several studies to justify the use of random over spatial cross-validation \citep[e.g.,][]{Cai2023, Xue2025}. 
Nevertheless, some of the results of \citet{Wadoux2021} align with previous research by \citet{Roberts2017} and \citet{Ploton2020}: random cross-validation applied to spatially clustered data results in considerable overestimation of map accuracy. What \citet{Wadoux2021} further demonstrate is that, depending on the spatial distribution of the data, spatial cross-validation might be overly pessimistic, at least for the two spatial cross-validation approaches they tested.
\citet{Mila2022} contribute to the debate by offering a different view on the design of spatial cross-validation: Rather than focusing solely on the training data, they argue that considering the prediction domain is key to achieving more reliable estimates of map accuracy (see section \ref{sec:adaptive_cv}).

In this paper, we seek to offer a perspective on key aspects of cross-validation for machine learning-based spatial mapping, an area where terminology, assumptions, and methodological choices are often sources of confusion. Due to their prevalence in current discussions and mapping studies, we mainly focus on cross-validation strategies. However, our findings are equally applicable to train/test splits, since the discussed resampling strategies can be applied to split the data into a train and a test subset in the same way. Furthermore, we focus on predictive modelling and acknowledge that modelling aiming at generating explanations might require different evaluation approaches.
To showcase challenges and considerations, we reproduce and extend the simulation study presented in \citet{Wadoux2021} for mapping forest above-ground biomass (AGB) for an area in the Amazon basin, based on \citet{Baccini2012}. AGB is predicted using a random forest model trained on 28 predictors, including climatic and remote-sensing variables. The original simulation study compared the performance of different evaluation methods for random and clustered sampling designs. We extend it to also include scenarios of moderately clustered sampling and spatial extrapolation (first row in Figure \ref{fig:pre_hoc}, see the appendix for full study details in reproducible form). Our goal is to synthesize existing knowledge on the evaluation of spatial predictions and to provide a forward-looking perspective by advocating a new class of evaluation methods that may help resolve the ongoing debate on cross-validation approaches.

\section{The extrapolation continuum}
\label{sec:extrapolation}
\begin{figure}
    \centering
    \includegraphics[width=1\linewidth]{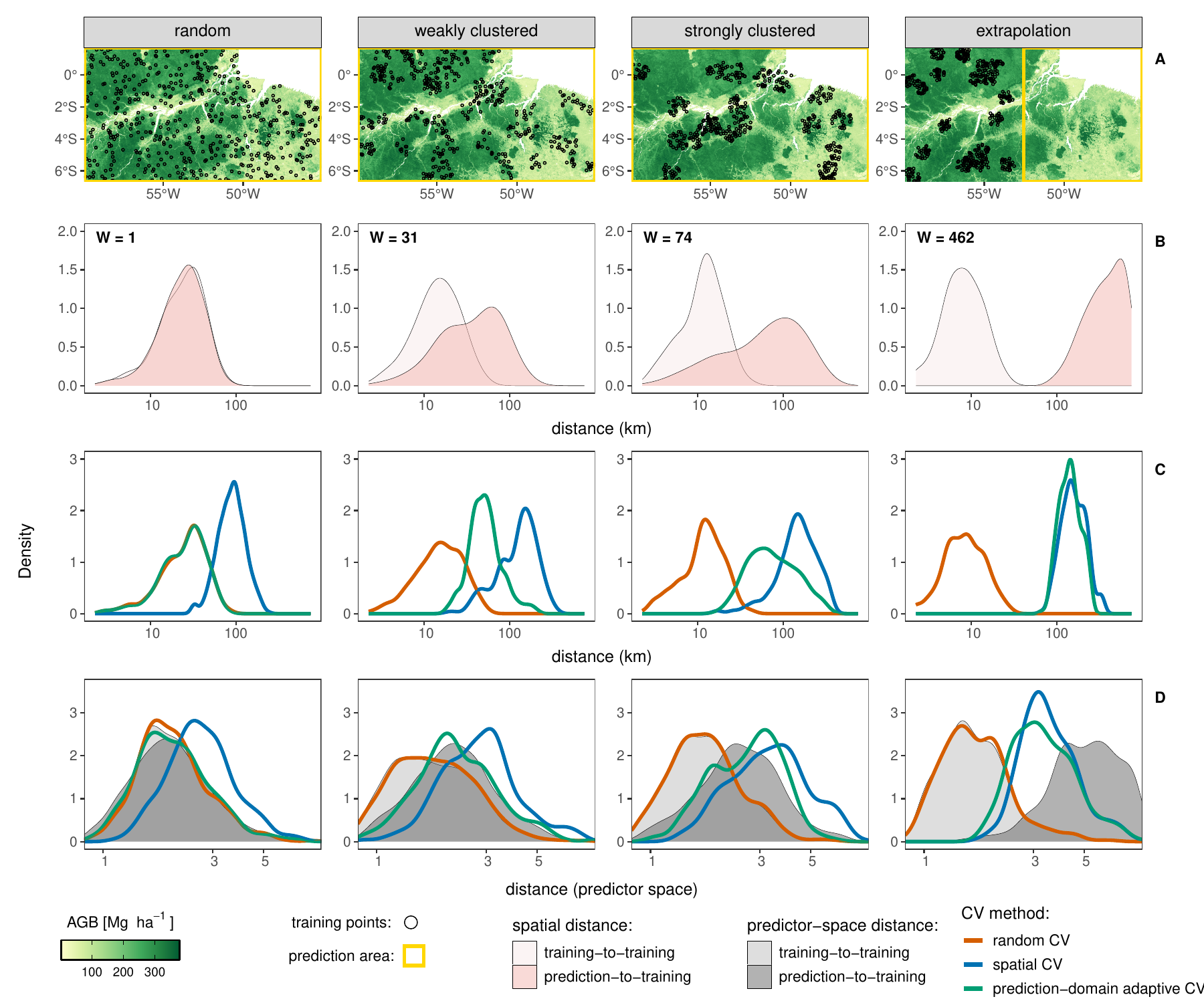}
    \caption{The upper row (A) shows sampling designs ranging from randomly distributed training point (left), through biased sampling where some areas are not covered by the training points, to clustered sampling and lastly to a setting requiring transfer to a new region (right). In the background, the above-ground biomass (AGB) is shown, while the black dots show the training points. 
    The second row (B) shows the distribution of spatial nearest neighbour distances between training points (\textit{training-to-training}) and between prediction locations and training points (\textit{prediction-to-training}) for the four prediction scenarios. The difference between the two density distributions is measured by the Wasserstein distance (bold W in the figure). A continuum from low spatial extrapolation (characterized by largely overlapping density distributions) on the left, towards strong extrapolation situations on the right (characterized by a large difference between the density distributions quantified by large W values), can be observed.
    The third row (C) shows the distribution of spatial nearest neighbour distances created when assigning random cross-validation folds (orange), spatial folds (blue) using spatial blocks for data splitting and prediction-domain adaptive folds calculated on geographic distances (green) based on k-Nearest Neighbour Distance Matching \citep[kNNDM, ][]{Linnenbrink2024}. The nearest-neighbour distance between folds was calculated as the distance between the validation points belonging to the held-out fold to the training points. 
    The last row (D) shows how these distances behave in the predictor space.
    For more details, refer to the appendix.}
    \label{fig:pre_hoc}
\end{figure}

As outlined above, previous research has shown that random cross-validation methods are suitable when the training points are randomly distributed in the prediction area, while spatial cross-validation is better suited towards extrapolation situations \citep[e.g.][]{Mila2022, deBruin2022, Linnenbrink2024}. 
In the literature on spatial predictive modeling, the terms "interpolation" and "extrapolation" are usually seen as two contrasting prediction scenarios \citep[e.g.,][]{Cressie1993, Roberts2017, Mila2022}. It is often argued that spatial extrapolation occurs when predicting to a completely new geographic area, while spatial interpolation describes scenarios where the training and prediction areas overlap (i.e., where the prediction points lie within the boundary box of the training points, see \citet{Cressie1993}). However, this binary definition misses the abundant cases where training and prediction areas overlap, but the training points are limited to few densely sampled areas. Such a clustered sampling results in gaps in the geographic space that are not covered by training points. Although, strictly speaking, we are not leaving the extent or convex hull of the training data, significant gaps have to be regarded similar to new areas and the prediction locations in these gaps cause challenges comparable to spatial extrapolation, while the prediction of locations close to the sampling locations would fall in the spatial interpolation category \citep[this has also been discussed in the general machine learning literature, see][]{Kuhn2013}. Hence, we argue that, in practice, extrapolation and interpolation do not form a clear binary distinction, but instead mark the opposite ends of a continuum (see second row in Figure \ref{fig:pre_hoc}). 

We translate this concept into a quantitative measure by calculating the difference between the density of spatial nearest neighbour distances between training points and the density of spatial nearest neighbour distances between prediction locations and training points. We argue that the nearest neighbour distance distribution between prediction locations and training points serves as a surrogate for the prediction situation. Then, the difference between these two density distributions, expressed as the Wasserstein statistic, quantifies the degree of extrapolation needed depending on the sampling design (W in Figure \ref{fig:pre_hoc}). For training points randomly distributed in the prediction area, the difference is small, indicating that the model is applied to grid cells that are nearby the training points and no extrapolation is required. With increasing clustering, more prediction locations are farther from the training points, and the degree of spatial extrapolation, quantified by the Wasserstein statistic, increases. The Wasserstein statistic is highest when the prediction domain is spatially distinct from the training domain, reflecting the full extrapolation case in our simulations (right column in Figure \ref{fig:pre_hoc}). 

Predictions become more challenging when moving from spatial interpolation to extrapolation. This is because spatial extrapolation is often associated with extrapolation in the predictor space, as both are typically linked through spatial autocorrelation. However, this relationship does not always hold. If important predictors are not included in the model, spatial extrapolation may not coincide with extrapolation in the available predictor space. Nevertheless, predictions in such settings remain challenging, as the model may be applied to regions that differ fundamentally from the training data, where different relationships may apply. At the same time, spatial extrapolation remains informative, particularly because models rely on available predictors, which may not fully represent the underlying drivers of the response variable.
In our simulation study, the distribution of nearest neighbour distances calculated as Euclidean distances between the scaled predictors at two locations follows largely those observed in the geographic space (Fig. \ref{fig:pre_hoc} row 4 compared to row 2). 
We believe that considering the degree of inter- or extrapolation is relevant in the context of data splitting for (cross-)validation as it will be outlined below.

\section{Extrapolation, independence and spatial autocorrelation in (cross-)validation: when and why does it matter?}
\label{sec:independence}

Spatial autocorrelation and the resulting spatial dependence structures in the response data are a central issue in the discussion around which of the different cross-validation schemes result in accurate estimates of map accuracy. Advocates for spatial resampling methods argue that independence between test and training points is needed to obtain reliable evaluation results \citep{Brenning2005, Roberts2017, Ploton2020}. Due to spatial autocorrelation, they note, response data that are geographically close to each other are also more similar to each other in the predictor space. The range of spatial autocorrelation present in the training points is often used to define spatial independence \citep{Roberts2017, Valavi2018, Ploton2020} where independence in this context is typically defined in a model-based manner, since it is based on the fitted variogram \citep{Brus1997}. Test points located within the autocorrelation range of the training points are then treated as not independent from the training data \citep{Roberts2017, Valavi2018, Ploton2020}. Thus, if a random cross-validation strategy is applied on clustered data, the test points are located near the training points and hence would be similar to them, violating independence between training and test data according to this definition. Such a random cross-validation scheme could not assess how well a model generalizes to new data, since it only tests how well it predicts to data that are nearby and hence typically similar to the training data. We show this in rows 3 and 4 of Figure \ref{fig:pre_hoc}: the nearest neighbour distances in both, spatial and predictor space, between random cross-validation folds (orange line) resemble the nearest neighbour distances between training points, but are much smaller than those between prediction locations and training points for all but the random sampling design where no spatial extrapolation is required. We consider the nearest neighbour distance distribution between prediction locations and training points as a surrogate for the prediction difficulty, and hence it comes with no surprise that random cross-validation often results in optimistically biased map accuracy estimations (Figure \ref{fig:post_hoc}). These results are consistent with numerous simulations and case studies, where random cross-validation consistently leads to an overestimation of map accuracy if the response data were spatially clustered \citep{Roberts2017, Wadoux2021, Mila2022, deBruin2022,Ludwig2023, Linnenbrink2024, Wang2025}. Furthermore, issues like overfitting cannot be detected in such a setting, since the spatial structures to which the model has overfitted are still present in the test data.
To solve these issues, spatial cross-validation methods have been proposed in which test and training points are separated by a given distance.
Many spatial cross-validation strategies to achieve independence have been developed and implemented in software packages. In R, the packages \emph{spatialsample} \citep{Mahoney2023} and \emph{mlr3spatiotempcv} \citep{Schratz2024} implement many spatial cross-validation approaches, including the widely used \emph{blockCV} \citep{Valavi2018} discussed in \citet{Roberts2017}. Comparable tools also exist for Python, such as \emph{spatial-kfold} \citep{Ghariani2023} and \emph{Verde} \citep{Uieda2018}.

On the other hand, \citet{Wadoux2021} argue against aiming for independence during cross-validation. They claim that aiming for independence leads to the creation of extrapolation situations during cross-validation, which then leads to overestimation of errors. This dilemma between aiming for spatial independence and the creation of extrapolation situations is not solvable from their point of view and argues against using spatial cross-validation methods. 
They demonstrated this with a simulation study -- reproduced and extended in this article -- which showed that spatial cross-validation severely overestimates the true mapping error for randomly or nearly randomly distributed training points. The distributions of nearest neighbour distances shown in Figure \ref{fig:pre_hoc} reflect this pattern: the distances between spatial cross-validation folds (blue line) are much longer than those encountered during predictions for all sampling designs except the spatial extrapolation scenario. These results have been confirmed in subsequent simulation studies, where spatial cross-validation yielded overly pessimistic estimates of map accuracy with the exception of strongly clustered sampling designs \citep{Mila2022, deBruin2022, Linnenbrink2024, Wang2025}.

\section{A call for prediction-domain adaptive evaluation}
\label{sec:adaptive_cv}

Recently, \citet{Mila2022}, \citet{Linnenbrink2024} and \citet{Wang2025} present alternative viewpoint on this discussion around independence between cross-validation folds. They suggest that enforcing independence between test and training points is not necessary and, instead, the cross-validation split should reflect the prediction situation which requires that the prediction area is taken into account when splitting the data.
In line with this perspective, \citet{Brenning2026} proposed a weighting scheme that approximates the prediction situation.
We consider these newly developed methods to belong to a new category of (cross-)validation methods, that we call \emph{prediction-domain adaptive evaluation}. While we describe their usage for cross-validation here, we emphasize that they can equally be used to split the data into training and test sets.

\begin{figure}
    \centering
    \includegraphics[width=1\linewidth]{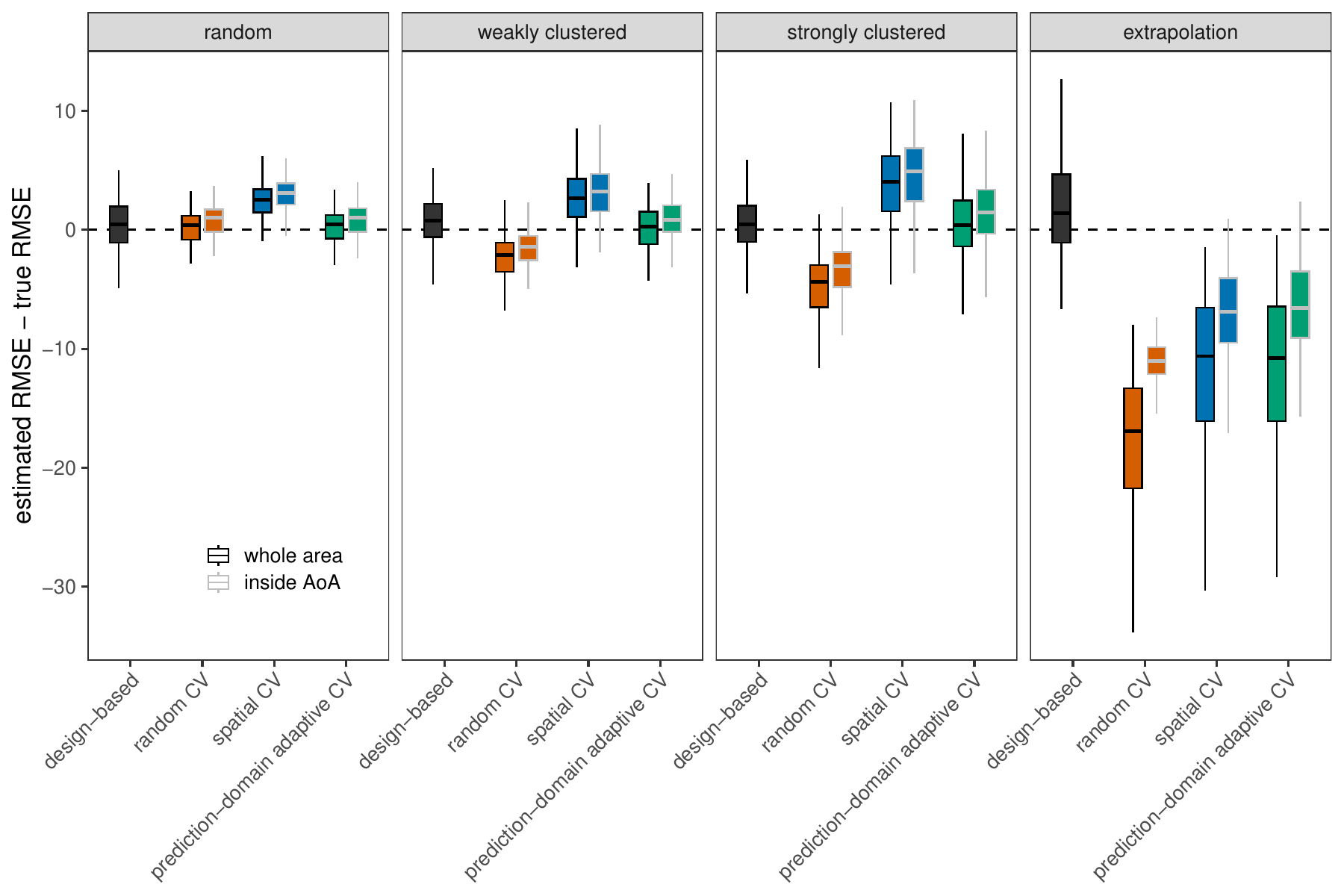}
    \caption{
    In our simulation study, we generated four training sampling designs (random, biased, clustered, and extrapolation) and trained random forest models to predict above-ground biomass (AGB) from 28 environmental predictors. For each sampling design, we tested a design-based evaluation, where independent test locations were randomly sampled from the entire prediction area and the three cross-validation strategies shown in Figure \ref{fig:pre_hoc} (random cross-validation, spatial cross-validation using spatial blocks, and kNNDM as a prediction-domain adaptive cross-validation). For each evaluation strategy we compared the estimated RMSE with the true RMSE calculated by comparing all predictions across the entire study area with their simulated truth. This process was repeated 100 times (for more details, refer to the appendix). In addition to the RMSE calculated for the whole prediction area (black outline colour), also the RMSE calculated only for the predictions that are inside the area of applicability (grey outline colour) is shown. The dashed horizontal line indicates zero difference, i.e., perfect estimation of the true RMSE (i.e., the closer the data are to the zero line, the more reliable the accuracy estimation strategy). Positive values mean overestimation of prediction errors by the evaluation strategy, negative values indicate underestimation.}
    \label{fig:post_hoc}
\end{figure}

The idea behind prediction-domain adaptive cross-validation, compared to random and spatial cross-validation, is visualized in the third row of Figure \ref{fig:pre_hoc}. Here, we used k-Nearest Neighbour Distance Matching \citep[kNNDM, ][]{Linnenbrink2024} as implemented in the R package CAST \citep{CAST} to represent the prediction-domain adaptive evaluation methods.
Prediction-domain adaptive cross-validation explicitly aims at resembling the prediction situation in terms of prediction difficulty as defined by the degree of extrapolation. Hence, prediction-domain adaptive cross-validation (green line) tries to produce nearest-neighbour distances that are comparable to those encountered during prediction, which can be done either in the geographical or the predictor space (here we focus on geographic space). For randomly distributed response data (left column in Figure \ref{fig:pre_hoc}), the prediction-domain adaptive cross-validation resembles the prediction situation by creating smaller distances between cross-validation folds. This comes close to random cross-validation (orange line), and artificially creating extrapolation situations as during spatial cross-validation (blue line) would result in more difficult prediction scenarios and hence typically overly pessimistic error estimates.
If the model is applied to situations requiring a larger degree of extrapolation, the prediction-domain adaptive cross-validation aims at creating large distances between cross-validation folds to resemble the large nearest neighbour distances between the prediction locations and the sampling locations. This is comparable to spatial cross-validation, but the aim is not to achieve independence, but rather matching the prediction situation. In this case, it is also not problematic that extrapolation situations are created during cross-validation, since this is exactly what the model encounters when applied to the prediction domain. In this way, the prediction-domain adaptive cross-validation solves the dilemma stated by \citet{Wadoux2021}, who argued that spatial cross-validation methods are caught between their pursuit of independence and the creation of extrapolation situation.
Hence, it is reasonable to assume that prediction-domain adaptive cross-validation adequately estimates the map accuracy if the prediction situation can be resembled.
This theoretical assumption is supported by our simulation study. Prediction-domain adaptive cross-validation consistently provided reliable proxies of map accuracy across all scenarios, except for the extrapolation scenario, which is discussed further below (Figure \ref{fig:post_hoc}).

\section{Where do the evaluation statistics hold?}

Building on the need for prediction-domain adaptive evaluation outlined above, it remains unclear where evaluation statistics can be expected to hold across the prediction domain. The evaluation strategies described above are commonly used to estimate the map accuracy of the prediction area assuming that, on average, the performance statistics apply to the prediction area. However, especially when making predictions for large or heterogeneous areas or even more obvious, in the case of significant spatial extrapolation, parts of the prediction area might be completely different compared to what has been encountered during model training and (cross-)validation. Machine learning models, however, typically fail when encountering predictor values that are different from what has been observed during training (extrapolation in predictor space). As a consequence, we cannot expect that the performance statistics hold for such areas because the model was not trained and validated for such conditions. This is reflected in Figure \ref{fig:post_hoc} especially when looking at the extrapolation case: even the prediction-domain adaptive or spatial cross-validation (where distances could be much better matched than for random cross-validation, see Figure \ref{fig:pre_hoc} bottom row) provide an overly optimistic estimate of the true map accuracy. Therefore, approaches to limit predictions and consequently the corresponding estimate of map accuracy to areas where the model was trained and validated are required. 
The area of applicability \citep{MeyerPebesma2021}, as one suggestion to achieve this, is based on a dissimilarity index where distances in the predictor space are used as a measure of dissimilarity. When using the dissimilarity observed during cross-validation as a threshold, the area of applicability is defined as the area where the model was enabled to learn about relationships and where, as a consequence, the cross-validation performance can serve as an estimate of map accuracy. Since the threshold is based on dissimilarities observed during cross-validation, this area is much smaller for random cross-validation than for the other cross-validation methods (see the appendix for more details). Hence, when using random cross-validation to estimate map accuracy, the accuracy is very high but this only applies to a limited area. When limiting the predictions to the area of applicability, Figure \ref{fig:post_hoc} shows that the cross-validation performance better fits the true map accuracy.

\section{Conclusion}
This paper contributes to the discussion around the suitability of different evaluation strategies in the field of spatial predictive mapping. Here, we did not address modelling for explanation, which likely requires different evaluation strategies that enforce extrapolation, such as blocking in predictor or geographic space. Instead, we aimed at providing a consolidated approach for evaluating predictive models: the prediction-domain adaptive evaluation. Our key messages are as follows.

\begin{enumerate}
    \item \textbf{Unbiased estimation of map accuracy:}
    An unbiased estimation of map accuracy requires a probability sample drawn from the prediction area. All other approaches, including test or cross-validation sets split randomly or spatially, only serve as proxies of the map accuracy.
    \item \textbf{Relevance of cross-validation:}
    When no probability sample is available, a separate test set, separated according to the same logic as cross-validation folds, should be favoured to estimate the map accuracy due to the risk of data leakage when using cross-validation for model tuning and final map accuracy estimation. However, a reliable cross-validation strategy remains indispensable during model development when selecting predictors, for tuning of hyperparameters, refining model structure, and to ultimately improve model quality.
    \item \textbf{Scope of different cross-validation strategies:}
    A wide range of non-spatial and spatial cross-validation strategies has been proposed, sparking an ongoing debate regarding their appropriate use. Random cross-validation is generally suitable when training points are randomly or systematically distributed across the prediction area, whereas spatial methods in general are preferable to assess extrapolation tasks. However, in practice, many use cases fall between these extremes, resulting in what can be described as an extrapolation continuum.
    \item \textbf{Towards prediction-domain adaptive evaluation:} We argue for a new category of evaluation methods that flexibly adapt to this extrapolation continuum: prediction-domain adaptive evaluation. Methods in this category offer a more balanced approach by adapting cross-validation according to the prediction domain. Two complementary approaches exist: aligning the resampling to the prediction difficulty, and weighting the cross-validation results based on the prediction difficulty. The difficulty of prediction can be measured both in the geographic space and in the predictor space. We argue that these methods provide more realistic estimates of map accuracy.
    \item \textbf{Limits of cross-validation:}
    Cross-validation-based accuracy estimates are only meaningful for new prediction situations that resemble the cross-validation folds in terms of prediction difficulty. When predictions extend into unsampled or substantially different regions (in terms of predictor properties) which cannot be resembled during cross-validation, performance statistics do not generalize reliably. For this reason, both predictions and corresponding accuracy assessments should be limited to regions that were represented in the training data (i.e. area of applicability).
\end{enumerate}

In summary, while acknowledging that cross-validation never replaces design based inference to estimate the map accuracy, both spatial and random evaluation approaches may be appropriate to estimate the map accuracy. Their suitability, however, depends on the prediction context. We believe that the proposed prediction-domain adaptive evaluation approach helps consolidate the ongoing discussion on various strategies by providing a balanced approach that yields more suitable estimates of map accuracy during cross-validation. This, in turn, supports model tuning and, in the absence of a probability sample, provides an estimate of the final map accuracy. 

\section*{Acknowledgements}

This project has received the financial support of the European Union’s Horizon Europe research and innovation programme under the Marie Skłodowska-Curie grant agreement No. 101147446, as well as the project BEyond within the priority Program 1374 "Biodiversity-Exploratories" (512284513) and the TRR 391 Spatio-temporal Statistics for the Transition of Energy and Transport (520388526), both funded by the Deutsche Forschungsgemeinschaft (DFG, German Research Foundation).

\section*{Conflict of Interest statement}
The authors declare no competing interests.

\section*{Author Contributions}
All authors were responsible for conceptualization. JL and JN were responsible for methodology development, software implementation, formal analysis, visualization and manuscript writing. HM contributed substantially to methodological refinement, interpretation of results, and manuscript writing. All authors reviewed and approved the final version of the manuscript.

\section*{Code and Data Availability}
The code and data to reproduce the simulations and to create the appendix are available on Github: \url{https://github.com/LOEK-RS/prediction-domain-adaptive-validation}.

\bibliographystyle{unsrtnat}
\bibliography{literature} 

\end{document}